# Spin dynamics in the geometrically frustrated multiferroic $CuCrO_2$

M. Poienar[1], F. Damay[2], C. Martin[1], J. Robert[2], and S. Petit[2]

[1] *Laboratoire CRISMAT, ENSICAEN, UMR 6508 CNRS, 6 Boulevard du Maréchal Juin, 14050 Caen Cedex, France*

[2] *Laboratoire Léon Brillouin, CEA-CNRS, UMR 12, CEA-Saclay, 91191 Gif-sur-Yvette Cedex, France*

Contact author : sylvain.petit@cea.fr

ABSTRACT

The spin dynamics of the geometrically frustrated triangular antiferromagnet multiferroic $CuCrO_2$ have been mapped out using inelastic neutron scattering. The relevant spin Hamiltonian parameters modelling the incommensurate modulated helicoid have been determined, and correspond to antiferromagnetic nearest and next-nearest neighbour interactions in the *ab* plane, with a strong planar anisotropy. The weakly dispersive excitation along *c* reflects the essentially two-dimensional character of the magnetic interactions and according to classical energy calculations it is weakly ferromagnetic. Our results clearly point out the relevance of the balance between a ferromagnetic coupling between adjacent planes and a weakly antiferromagnetic next-nearest neighbour interaction in stabilising the three-dimensional ferroelectric magnetic order in $CuCrO_2$. This novel insight on the interplay between magnetic interactions in $CuCrO_2$ should provide a useful basis in the design of new delafossite-based multiferroic materials.

INTRODUCTION

For more than a decade, the study of frustrated antiferromagnets has been a fascinating subject of condensed matter physics, as the macroscopic degeneracy of the classical ground state of these systems is considered to lay the grounds for challenging novel physics. The perfect triangular lattice, an archetype for the study of geometric frustration in two dimensions (2D), has recently attracted much attention, owing to the discovery of multiferroic properties in the triangular systems delafossite oxides $CuFeO_2$ [1] and $CuCrO_2$ [2]. As shown on Figure 1a, these compounds are characterised at room temperature by a stacking of perfect triangular arrays as their $R\bar{3}m$ symmetry ensures the isotropy of the in-plane couplings. Despite their inherent frustration, these systems find a way to lift their macroscopic degeneracy and to achieve a complex three-dimensional (3D) magnetic ordering below a Néel ordering temperature $T_N$. In the case of $CuFeO_2$ a transition from the $R\bar{3}m$ to $C2/m$ symmetry is found to accompany the magnetic ordering ; this distortion is believed to help lifting the degeneracy of the frustrated magnetic lattice, to achieve an Ising-like 4-sublattice (↑↑↓↓) antiferromagnetic (AF) order at low temperature [3], [4], [5]. In $CuCrO_2$, an incommensurate magnetic structure, derived from the classical 120° spin configuration expected for a perfect planar triangular antiferromagnet [6], has been reported [7], [8]. The transition to the non-collinear magnetic state in this compound coincides with a spontaneous ferroelectric polarisation, evidencing the intimate coupling between the magnetic and electric order parameters [9]. From a microscopic point of view, the origin of this coupling remains quite puzzling, as the trigonal symmetry of these compounds imposes severe constraints on the standard theoretical models. Nonetheless, Arima *et al.* [10] propose that the electric polarisation may actually result from a subtle modulation - a consequence of spin-orbit coupling - of the hybridisation between the *3d* ions carrying the spin and the bonding oxygen ions.

In this article, we focus on the case of multiferroic delafossite $CuCrO_2$. Unlike $CuFeO_2$, no symmetry lowering has ever been evidenced in this compound yet. Owing to its layered topology, significant antiferromagnetic exchange coupling is expected within the plane ($J_{ab} > 0$), though the influence of next-neighbour coupling $J_{NN}$ (which has been shown to be prominent in $CuFeO_2$ [11]) remains to be assessed (Figure 1a). In contrast, the rather short correlation length along the *c*-axis revealed by neutron experiments supports the highly 2D character of the compound, and thus a weak magnetic exchange between the planes ($J_c$) is expected. Amongst the two models proposed in [8] (Figures 1b and 1c) to describe the magnetic spin configuration of $CuCrO_2$, the helicoidal model is the only one that can drive, in the framework of the mechanism proposed by Arima *et al*. [10], the experimentally observed ferroelectricity within the hexagonal plane [12]. The helicoidal structure of $CuCrO_2$ has actually been confirmed recently by Soda et al. [13], using polarised neutron scattering.

In order to study the magnetic interactions ($J_{ab}$, $J_{NN}$, $J_c$) we carried out inelastic neutron scattering experiments on a single crystal of $CuCrO_2$. In parallel, we performed energy calculations based on a standard Heisenberg model to determine the ($J_{NN}$, $J_c$) phase diagram of the classical ground state of $CuCrO_2$. We find that the INS data can be parameterised by a spin-wave model taking into account an easy-plane anisotropy term and a strong in-plane AF exchange coupling $J_{ab}$. We emphasise the excellent agreement obtained between the experimental data and the spin dynamics modelling based on a helicoidal structure. Our results show that a ferromagnetic inter-plane coupling $J_c$ can explain the subtle incommensurate deviation away from the classical 120° structure observed in $CuCrO_2$. In addition, a weak next-nearest neighbour coupling $J_{NN}$ is evidenced for the first time in this compound.

EXPERIMENTAL METHODS

A single crystal of CuCrO$_2$ (70 mg) was grown by the flux technique, following [14]. A polycrystalline sample (5 g) was also prepared according to [8]. Inelastic neutron scattering experiments were performed on the thermal (2T) and cold (4F2) neutron triple-axis spectrometers at Laboratoire Léon Brillouin (LLB)-Orphée (Saclay, France). Because of the challengingly small volume of the sample, focusing monochromators and analyzers were used to optimise the intensity, as well as standard $k_f$ values ($k_f$ = 2.662Å$^{-1}$ or 1.550Å$^{-1}$) depending on the desired resolution. Higher order contaminations were removed with pyrolytic graphite or nitrogen-cooled Be filter placed in the scattered beam. The sample was mounted on an aluminium holder and aligned in the (HHL) scattering plane.

RESULTS

Based on a simple analysis of the magnetic structure, the properties of CuCrO$_2$ should be reasonably well described by the following spin Hamiltonian :

$$H = \sum_{i,j}^{\substack{in \\ plane}} J_{ij} S_i S_j + J_c \sum_{i,j}^{\substack{inter \\ plane}} S_i S_j \quad (1)$$

$J_{ij}$ are the planar exchange couplings between spins located at sites $i$ and $j$ on a triangular plane and correspond to nearest ($J_{ab}$), next-nearest ($J_{NN}$) and next-next-nearest ($J_{NNN}$) neighbour interactions within this triangular plane. $J_c$ denotes the coupling between nearest neighbours in adjacent layers. Aiming at a more physical understanding of the model, we have calculated the classical energy, assuming that the spins order with a propagation vector of the form **k** = ($q$ $q$ 0), as experimentally observed. To this end, the ENERMAG program [15] was used to generate a magnetic phase diagram as a function of the parameters $J_{NN}/J_{ab}$ and $J_c/J_{ab}$. (Figure 2). The program explores, for each point in the J-space, the $q$ value for which the energy is minimum, using the generalisation of the Villain-Yoshimori theorem developed by Lyons, Kaplan and Freiser [16], [17]. Figure 2 shows that if $J_c$ = 0, the propagation vector is **k** = (⅓ ⅓ 0), which corresponds to the classical commensurate 120° structure. If $J_c$ ≠ 0, an incommensurate structure will be the most stable, except in the case of a strong antiferromagnetic $J_{NN}$ coupling ($J_{NN}/J_{ab}$ >> 0), which stabilises a collinear phase with **k** = (½ ½ 0). Interestingly, the commensurate 120° pattern vanishes as soon as a small $J_c$ coupling is introduced, and an incommensurable phase with $q$ close to ⅓ is stabilised. To further understand this effect of $J_c$, we have calculated analytically the classical energy corresponding to (1), for the same **k** = ($q$ $q$ 0) propagation vector. The different contributions associated with subsequent neighbours are listed hereafter (2):

$$E_{ab} = J_{ab}S(\cos 4\pi q + 2 \cos 2\pi q)$$

$$E_{NN} = J_{NN}S(1 + 2 \cos 6\pi q)$$

$$E_c = J_cS(1 + 2 \cos 2\pi q)$$

The most simple approach is to neglect in the first place the $J_{NN}$ interaction, the total energy is therefore $E = E_{ab} + E_c$. Minimising $E$ with respect to $q$ gives :

$$\frac{J_c}{J_{ab}} = -(1 + 2 \cos 2\pi q) \qquad (3)$$

in agreement with the ENERMAG numerical calculation. Because the $q$ value depends on the sign of $J_c/J_{ab}$, knowing that $q$ = 0.329 from neutron diffraction, this simple model leads to the conclusion that a ferromagnetic $J_c/J_{ab}$ =-0.05 can account for the incommensurate structure.

More generally, taking $J_{NN}$ into account and minimising the classical energy $E$ with respect to $q$ leads to the following relation between $q$ ($\neq 0.5$) and $J_{NN}/J_{ab}$ and $J_c/J_{ab}$ in the phase diagram:

$$\frac{J_c}{J_{ab}} = \frac{J_{NN}}{J_{ab}}(3 - 12\cos^2 2\pi q) - (1 + 2\cos 2\pi q) \qquad (4)$$

as illustrated on Figure 2 for $q = 0.329$. Interestingly, weakly ferromagnetically coupled planes ($0 > J_c/J_{ab} > -0.05$) require only relatively weak and antiferromagnetic next-neighbour coupling to stabilise the $q = 0.329$ structure. On the other hand, an increase in the ferromagnetic coupling between planes $J_c$ with respect to the critical value $J_c/J_{ab} = -0.05$ will impose a change of sign of the next-neighbour coupling from antiferromagnetic to ferromagnetic in order for the $q = 0.329$ structure to stay stable. On a more general point of view, the effect of $J_{NN}$ is more critical to the incommensurability for larger $|J_c|$.

The INS measurements carried out on the powder sample give an overview of the excitation spectrum and of its temperature evolution. Figure 3 shows the neutron intensity as a function of energy and momentum transfer at 5K and 40K. It reveals below $T_N$ a magnetic excitation spectrum located between 1 and 9 meV, characterised by a broad maximum at ~5 meV. Because of the experimental resolution (1 meV), we cannot assert the existence of a spin gap at the $Q$ position of the magnetic Bragg peak ($q\ q\ 0$). The above features can be simply explained by the existence of a planar anisotropy term in the spin Hamiltonian : with this additional term, a global rotation of the spins in the easy-plane still does not cost any energy, by contrast to an out-of-plane rotation, which will cost some anisotropy energy. The spin dynamics spectrum will have accordingly one branch having zero energy at the magnetic Bragg point (the Goldstone mode of the structure), together with other branches exhibiting a gap representative of the easy-plane anisotropy energy. In the powder average spectrum, this leads to significant scattering down to zero energy, as well as to a maximum at the energy of the gap. At T = 40K, well above $T_N$, the spectral weight shifts to lower energies, and significant magnetic scattering is still measured, reminiscent of short range magnetic fluctuations.

To go further in the analysis, we carried out triple-axis experiments on a single-crystal of $CuCrO_2$, in the energy range previously determined E < 9 meV. We mapped out the magnetic excitations propagating within the hexagonal plane, along the [H H 0] direction, at 10K (Figure 4) for $k_f$ = 1.550Å$^{-1}$ and $k_f$ = 1.97Å$^{-1}$ (because of the small volume of sample available, a larger $k_f$ was employed to increase the intensity to the detriment of the instrumental resolution). The spectrum (Figure 4a) can be described as follows : a main branch stems from the magnetic Bragg point (*q q* 0) with *q* ~ 0.33. This branch does not show any clear energy gap within instrumental resolution, as is also seen on energy scans along [⅓ ⅓ 0] (not shown). However, a linear fit to the dispersion shows that if there is a gap it is actually smaller than 0.6 meV (Figure 4a and 4b). This branch describes an arch whose maximum reaches about 8 meV at the zone boundary (½ ½ 0) (upper red dotted line on Figure 4c). The second observable feature is weaker and is only clearly detected at high momentum, close to the zone boundary (½ ½ 0) (Figure 4c). It appears around 4 meV and is relatively flat compared to the previous excitation, but can be clearly identified on the energy scan taken at constant Q = [0.425 0.425 0] (arrow on Figure 4d). A third feature around the zone centre (⅓ ⅓ 0) (shown by an arrow on Figure 4c), is reminiscent of the 4 meV gap already evidenced on the powder average dispersion.

To model the spin dynamics, spin-wave calculations were performed using the *Spinwave* software developed at LLB. Based on the Holstein-Primakov approximation, the code diagonalises any spin Hamiltonian based on equation (5), which is derived from equation (1) :

$$H = J_{ab} \underbrace{\sum_{i,j}}_{\substack{in \\ plane}} S_i S_j + J_{NN} \underbrace{\sum_{i,j}}_{\substack{in \\ plane}} S_i S_j + J_c \underbrace{\sum_{k,l}}_{\substack{inter \\ plane}} S_k S_l + D \sum_i (S_i \cdot n_i)^2 \qquad (5)$$

It takes into account (isotropic or anisotropic) exchange couplings acting between neighbouring spins, as well as single ions anisotropy terms modelled by $D \sum_i (S_i \cdot n_i)^2$. By definition, if $D$ is positive, $n$ denotes the vector perpendicular to the easy-plane anisotropy. If $D$ is negative, it accounts then for an easy-axis anisotropy, with $n$ being in this case the easy-axis direction. Once the spin-wave energies are known, spin correlations functions are calculated to obtain the dynamical structure factor observed by inelastic neutron scattering experiments. Interactions are limited to next-nearest-neighbour. As the incommensurate deviation from $q = ⅓$ is small, the calculations were made with a simplified perfect 120° spin pattern in a commensurate magnetic cell ($3a$, $3a$, $c$) containing 27 magnetic atoms; the easy plane, and relative spin orientations were determined according to [8]. Two models were actually considered for the helicoidal magnetic structure : the first one considers that the spin rotation envelope is circular, that is that the spin value S = 1.5 is not modulated from one Cr site to the other. The second one takes into account the elliptic spin modulation described in [8], in which the spin amplitude depends on its orientation with respect to the $c$ axis.

Figure 5 and 6 illustrate the dynamical structure factors calculated for the non-modulated and modulated helicoidal models, respectively, as a function of the wave-vector [H H 0] and of the energy transfer E. In the case of the non-modulated model, two main modes are observed, denoted ($\alpha$) and ($\beta$) on Figure 5. On the experimental data, the $\beta$ mode shows a 4 meV gap and flattens out quite quickly close to the zone boundary (½ ½ 0) : to describe it correctly, it is necessary to add in the modelling a small next-nearest neighbour coupling $J_{NN}$, as can be seen by comparing the dispersions calculated without (Figure 5a) and with antiferromagnetic next-nearest neighbour coupling ($J_{NN}$ = 0.25 meV) (Figure 5b). As is illustrated on Figures 5c and 5d, the experimental data is accordingly best reproduced with $J_{ab}$ = 2.30 meV, $J_{NN}$ = 0.25 meV and $D_{(100)}$ = 0.40 meV. The influence of inter-layer coupling $J_c$ could not be evidenced in the modelling and all the calculations were performed with $J_c$ = 0.

In the second model, which corresponds to a modulated helicoidal structure, the elliptical modulation of the spin rotation envelope was reproduced using a spin moment $S$ = 1.4 (spin parallel to [0 0 1]) and two $S$ = 1.1 (spin oriented at 120° and 240° with respect to [0 0 1]). This significantly affects the spin-wave dispersion : in addition to the ($\alpha_m$) and ($\gamma_m$) modes, another branch ($\beta_m$), stemming from the magnetic Bragg point, is clearly observed (Figure 6a). In this case as well, it is necessary to add in the modelling a small next-nearest neighbour coupling $J_{NN}$, to account for the softening of $\beta_m$ close to the zone boundary (½ ½ 0) (Figure 6b). The experimental data is here best reproduced with $J_{ab}$ = 2.30 meV, $J_{NN}$ = 0.12 meV and $D_{(100)}$ = 0.40 meV, and keeping $J_c$ = 0 (Figures 6c and 6d).

Scattering profiles along the [H H L] directions are illustrated on Figure 7, which shows data taken at fixed energy E = 2.5 meV, as a function of H, for a number of L values. We observe that the peak positions are only weakly L-dependent, demonstrating that the coupling $J_c$ between the planes is extremely weak. This result is fully consistent with the observation of very broad magnetic Bragg peaks at 1.5K [8], which was interpreted as short range (~ 200Å) magnetic correlations along *c*, resulting from a weak inter-plane coupling. Examples of the profiles calculated for different values of $J_c$ are illustrated in Figure 7b : it is not possible to determine the sign of $J_c$ or estimate its value within the accuracy of the results, however, we nevertheless can approximate the bandwidth of the dispersion along the *c*-direction in the range [-0.2, 0.2] meV, which corresponds to -0.1 < $J_c/J_{ab}$ < 0.1 in the phase diagram of Figure 2.

The modelling parameters obtained from the inelastic scattering study can now be compared with the theoretical calculations summarised in Figure 2. Even though we are unable to determine the nature or strength of the inter-plane coupling $J_c$ from the experimental dispersion along $c$, inelastic scattering data in the $ab$ plane clearly indicate that next-nearest neighbour coupling $J_{NN}$ is not negligible in $CuCrO_2$ : it is antiferromagnetic, and can be estimated to range between [0.1-0.25] meV depending on the chosen model. This is quite an unforeseeable result, which only inelastic scattering could evidence. Using $J_{ab}$ = 2.30 meV from the INS modelling of the magnetic structure, this leads, for the non-modulated (modulated) models to $J_{NN}/J_{ab}$ ~ 0.108 (0.052), which in turn, using equation (4) and $q$ = 0.329, gives an estimate of $J_c/J_{ab}$ ~ -0.017 (-0.033) (illustrated as an empty (filled) star on the phase diagram of Figure 2). In absolute value, this is only slightly less than the $J_c/J_{ab}$ value estimated for $J_{NN}$ = 0 ($J_c/J_{ab}$ ~ -0.05 in that case), and emphasises the fact that, whatever the chosen model, the helicoidal magnetic structure of $CuCrO_2$ is stabilised by a weak antiferromagnetic next-nearest neighbour coupling ($J_{NN}$ = 0.25 (0.12) meV), in addition to a weak *ferromagnetic* inter-plane coupling ($J_c$ ~ -0.04 (-0.08) meV). The corresponding dispersion, calculated in the case of the modulated helicoid model for $J_{ab}$ = 2.30 meV, $J_c$ = -0.08 meV, $D_{(100)}$ = 0.40 meV, is illustrated in Figure 7c and is in good agreement with the data.

With the available experimental data, it is actually quite difficult to decide which model describes $CuCrO_2$ best. The modulated helicoid should provide in theory the closest description, but it does not reproduce exactly the softening of the spin-wave mode around (½ ½ 0), and probably leads to an underestimation of $J_{NN}$ in the calculation. Both models still remain fairly simple, as they do not take into account the magnetic incommensurability, but both also give consistent values of the magnetic couplings, whose main characteristic is to lie very close to the limit of stability of the non-collinear magnetic phase.

DISCUSSION

In $CuCrO_2$, calculations and experimental observations have thus highlighted the delicate balance between $J_{NN}$ and $J_c$ to stabilise the incommensurate helicoidal magnetic order. An improved model is however required to further understand some of the more subtle features of the inelastic scattering data. Very few studies of the spin dynamics in other transition metal (TM) delafossite compounds are actually available in the literature and they mainly concern undoped and Al-doped $CuFeO_2$. Ye *et al.* in [11] describe the spin dynamics in the collinear 4-sublattice structure of $CuFeO_2$ using a $J_{ab}$ ~1.14 meV, a strong $J_{NN}$ ~ 0.50 meV, and an additional next-next-nearest neighbour coupling $J_{NNN}$ ~ 0.65 meV, though the authors themselves acknowledge that there are no really clear physical grounds to it. In $CuFeO_2$, the excitation along *c* is also clearly dispersive, and the inter-plane coupling is accordingly strongly antiferromagnetic ($J_c$ ~ 0.33 meV). A quantitative analysis of the magnetic couplings in $CuFeO_2$ is beyond the scope of this paper, but we note that, in a fairly broad interpretation of our ENERMAG calculations, larger coupling values will lead to a larger deviation from commensurability, and for a strong $J_{NN}$ coupling, to a collinear phase, in agreement with what is observed for $CuFeO_2$. A more detailed INS investigation of other TM delafossites is required at this point to draw a general picture, but the relative strengths of the magnetic paths $J_{ab}$, $J_{NN}$ and $J_c$ dictated by the structural topology are certainly strongly dependent on the TM element. The strength of the inter-plane coupling $J_c$, intermediated by the linear copper bonding, seems, in particular, to depends highly on the nature of the TM cation.

Ye *et al.* also point out that, even if the collinear structure of $CuFeO_2$ shows magnetic Bragg reflections at $(\frac{1}{4}\ \frac{1}{4}\ \frac{3}{2})$, the dispersion exhibits two minima at the $Q$ positions corresponding to the two incommensurate magnetic Bragg peaks positions. Indeed, the main difference between the spin dynamics in the collinear phase of $CuFeO_2$ and the incommensurate phase of $CuFe_{1-x}Al_xO_2$ (x = 0.0155) lies in the low energy part of the spectrum, where it can be seen that the energy of the spin-wave modes at these particular $Q$ vectors softens down to zero [18]. The spin dynamics in both structures is thus very similar, and it is plausible that the spin Hamiltonian remains basically identical. Based on these considerations, it seems that the transition towards the collinear 4-sublattice magnetic phase in $CuFeO_2$ is not purely an effect of magnetic coupling, but is rather related to an additional interaction, likely to be a strong coupling to the lattice in this case. In this sense, the substantial $J_{NN}$ value, coupled to a significant spin-lattice effect, can be understood as the driving force triggering the symmetry lowering transition in $CuFeO_2$. In $CuCrO_2$, although a $J_{NN}$ coupling has been evidenced, it has a much smaller amplitude than in $CuFeO_2$. In addition, the only sign of magneto-elastic coupling is the subtle relaxation of the compression along $c$ of the $CrO_6$ octahedron [8] at the onset of spin ordering, and accordingly, any actual symmetry lowering at $T_N$ remains hypothetical.

Interestingly, the stabilisation of the 4-sublattice collinear magnetic structure to the detriment of the incommensurate one in $CuFeO_2$ coincides with the disappearance of the multiferroic behaviour. In contrast, $CuCrO_2$, with weak $J_{NN}$ and $J_c$, remains down to the lowest temperatures a good example of an almost perfect 2D Heisenberg system on a triangular lattice. The transition towards a 3D magnetically ordered state at finite temperature, and concomitantly with a multiferroic phase, is only ensured by the small perturbations $J_{NN}$ and $J_c$.

CONCLUSION

In summary, the spin dynamics of the geometrically frustrated triangular antiferromagnet multiferroic $CuCrO_2$ have been mapped out using inelastic neutron scattering. We have determined the relevant spin Hamiltonian parameters, showing that the helicoidal model with a strong planar anisotropy correctly describes the spin dynamics. The weakly dispersive excitation along *c* reflects the essentially 2D character of the magnetic interactions, but the spin dynamics in $CuCrO_2$ clearly point out the relevance of the balance between a ferromagnetic coupling between adjacent planes and a weakly antiferromagnetic next-nearest neighbour interaction, to stabilize the 3D magnetic order.

Acknowledgements

Financial support for this work was partially provided by the French Agence Nationale de la Recherche, Grant No ANR-08-BLAN-0005-01.

REFERENCES

[1]    T. Kimura, J. C. Lashley, and A. P. Ramirez, *Physical Review B* **73**, 220401 (2006).

[2]    S. Seki, Y. Onose, and Y. Tokura, *Physical Review Letters* **101**, 067204 (2008).

[3]    N. Terada, S. Mitsuda, H. Ohsumi, and K. Tajima, *Journal Of The Physical Society Of Japan* **75**, 023602 (2006).

[4]    S. Mitsuda, N. Kasahara, T. Uno, and M. Mase, *Journal Of The Physical Society Of Japan* **67**, 4026 (1998).

[5]    F. Ye, Y. Ren, Q. Huang, J. A. Fernandez-Baca, P. C. Dai, J. W. Lynn, and T. Kimura, *Physical Review B* **73**, 220404 (2006).

[6]    M. F. Collins and O. A. Petrenko, *Canadian Journal Of Physics* **75**, 605 (1997).

[7]    H. Kadowaki, H. Kikuchi, and Y. Ajiro, *Journal Of Physics-Condensed Matter* **2**, 4485 (1990).

[8]    M. Poienar, F. Damay, C. Martin, V. Hardy, A. Maignan, and G. André, *Physical Review B* **79**, 014412 (2009).

[9]    K. Kimura, H. Nakamura, S. Kimura, M. Hagiwara, and T. Kimura, *Physical Review Letters* **103**, 107201 (2009).

[10]   T. H. Arima, *Journal Of The Physical Society Of Japan* **76**, 073702 (2007).

[11]   F. Ye, J. A. Fernandez-Baca, R. S. Fishman, Y. Ren, H. J. Kang, Y. Qiu, and T. Kimura, *Physical Review Letters* **99**, 157201 (2007).

[12]   K. Kimura, H. Nakamura, K. Ohgushi, and T. Kimura, *Physical Review B (Condensed Matter and Materials Physics)* **78**, 140401 (2008).

[13]   M. Soda, K. Kimura, T. Kimura, M. Matsuura, and K. Hirota, *Journal Of The Physical Society Of Japan* **78**, 124703 (2009).

[14]   M. Poienar et al., *in preparation* (2009).

[15] N. El Khayati, R. C. El Moursli, J. Rodriguez-Carvajal, G. Andre, N. Blanchard, F. Bouree, G. Collin, and T. Roisnel, *European Physical Journal B* **22**, 429 (2001).

[16] D. H. Lyons and T. A. Kaplan, *Physical Review* **120**, 1580 (1960).

[17] M. J. Freiser, *Physical Review* **123**, 2003 (1961).

[18] N. Terada, S. Mitsuda, T. Fujii, and D. Petitgrand, *Journal Of Physics-Condensed Matter* **19**, 145241 (2007).

FIGURE CAPTIONS

figure 1 : Modelling of the magnetic structure of $CuCrO_2$ : (a) Illustration of the nearest-neighbour ($J_{ab}$) and next-nearest neighbour ($J_{NN}$) magnetic interactions within the triangular plane, and between the planes ($J_c$) in the delafossite structure. Each $Cr^{+3}$ is surrounded by 6 nearest neighbours and 6 next-nearest neighbours within the plane, and by 2 x 3 neighbours in the adjacent planes above and below. Possible magnetic structures of $CuCrO_2$ : modulated helicoid with the spin rotation axis parallel to [110] (b) and cycloid with the spin rotation axis parallel to [1$\bar{1}$0] (c). The propagation vector is shown as a full arrow and corresponds to **k** = ($q$ $q$ 0) ($q \sim 0.329$).

Figure 2 : Magnetic phase diagram in the ($J_c$, $J_{NN}$) plane of a triangular lattice with strong in-plane antiferromagnetic coupling $J_{ab}$. $J_c$ and $J_{NN}$ are inter-plane exchange and in-plane next-nearest neighbour couplings, respectively. $q$ describes the value of the propagation vector **k** = ($q$ $q$ 0). Stars refer to the coupling values determined in the case of the non-modulated (empty symbol) and modulated (filled symbol) helicoidal magnetic structures.

Figure 3 : Inelastic neutron scattering powder spectra recorded at 5K and 40K ($k_f$ = 2.662Å$^{-1}$). On the 5K spectrum is also shown the corresponding neutron elastic scattering data (from [8]).

Figure 4 : Inelastic neutron scattering spectra at 10K along [H H 0] ($k_f$ = 1.550Å$^{-1}$). (a) Spin-wave dispersion (intensity contour). Red lines are a fit to the dispersion. (b) Representative spin-wave excitation scans at constant energy transfer (E = 0.9, 2.5 and 5.6 meV). Data are shifted for clarity. (c) Spin-wave dispersion (intensity contour) ($k_f$ = 1.970Å$^{-1}$). (d) Constant

wave-vector scans along [0.425 0.425 0] (dashed line on (c)) with $k_f$ = 1.970Å$^{-1}$ at 10K. Arrows underline features described in the text.

Figure 5 : Calculated dynamical structure factor of CuCrO$_2$, plotted versus wave vector [H H 0] and energy transfer E, using the Heisenberg Hamiltonian (1) in the case of the non-modulated helicoidal model (a) without $J_{NN}$ and (b) with an additional $J_{NN}$ coupling term (see text). The different spin-wave modes are outlined in white. (c) and (d) Superimposition of the calculated model (dark grey lines) on the experimental inelastic scattering data map at 10K recorded at $k_f$ = 1.550Å$^{-1}$ and $k_f$ = 1.970Å$^{-1}$ respectively.

Figure 6 : Calculated dynamical structure factor of CuCrO$_2$, plotted versus wave vector [H H 0] and energy transfer E, using the Heisenberg Hamiltonian (1) in the case of the modulated helicoidal magnetic structure model (a) without $J_{NN}$ and (b) with an additional $J_{NN}$ coupling term (see text). The different spin-wave modes are outlined in white. (c) and (d) Superimposition of the calculated model (dark grey lines) on the experimental inelastic scattering data map at 10K recorded at $k_f$ = 1.550Å$^{-1}$ and $k_f$ = 1.970Å$^{-1}$ respectively.

Figure 7 : Inelastic neutron scattering study along [H H L] ($k_f$ = 1.550Å$^{-1}$) : (a) Spin-wave excitation scans at constant energy transfer (E = 2.5 meV) along the [H H L] directions (L = 0, 0.35, 0.5, 0.7, 1). Data are shifted for clarity. (b) Comparison between experimental and modelled dispersion profile along [H H 0] calculated for different values of interplanar coupling $J_c$ = 0, 0.2 and -0.2 meV in the modulated helicoid case. (c) Intensity contour of the calculated dispersion in the [H H L] plane (E = 2.5 meV), for a ferromagnetic inter-planar coupling $J_c$ = -0.08 meV in the modulated helicoidal structure case.

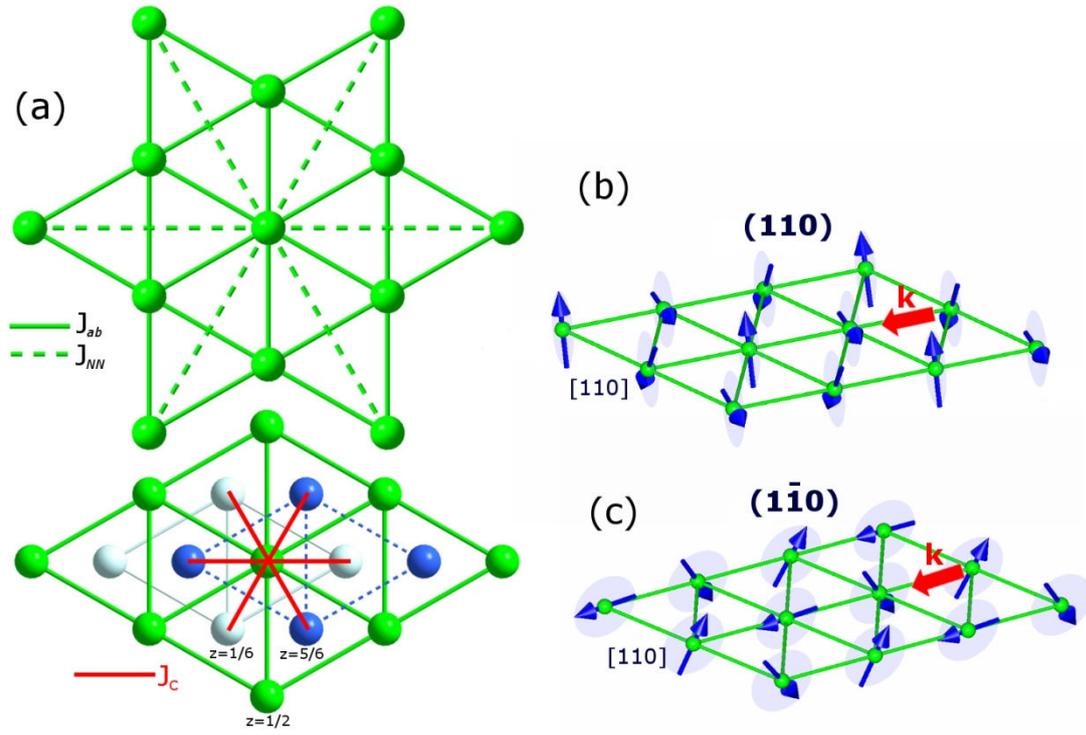

Figure 2

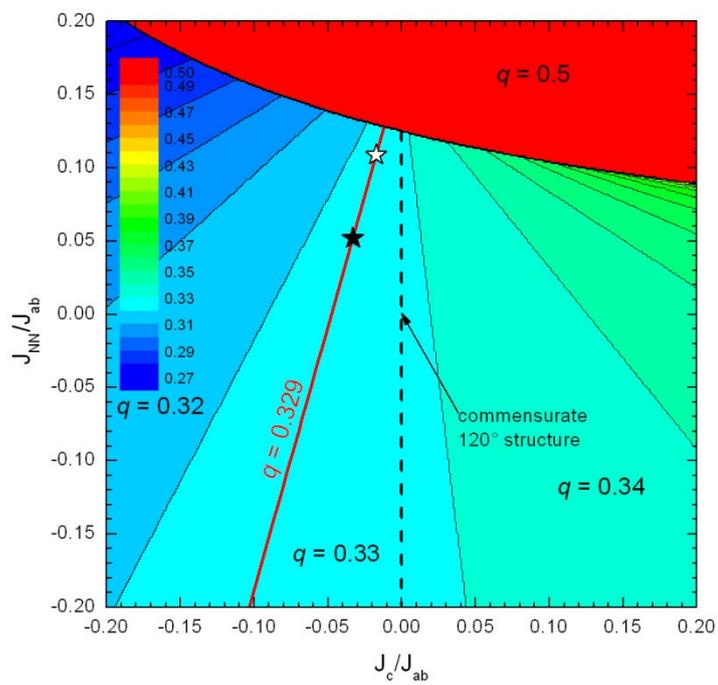

Figure 3

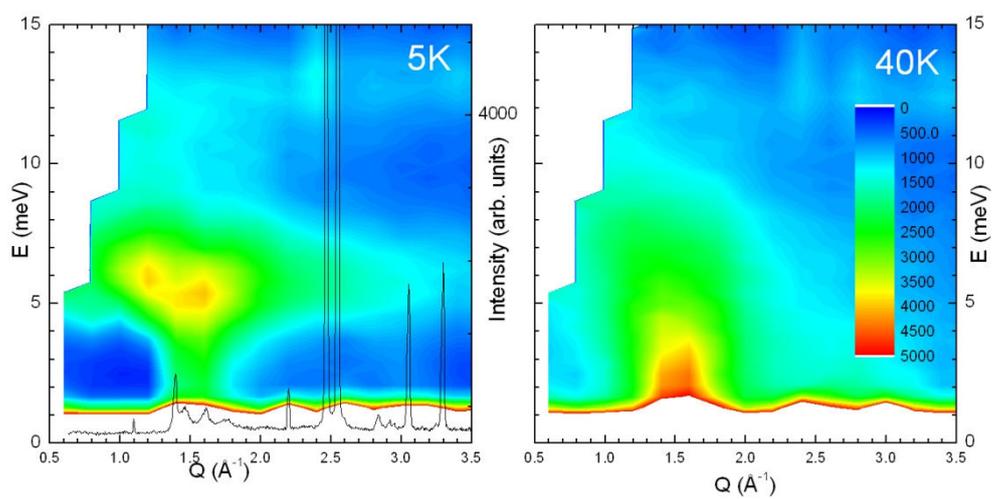

Figure 4

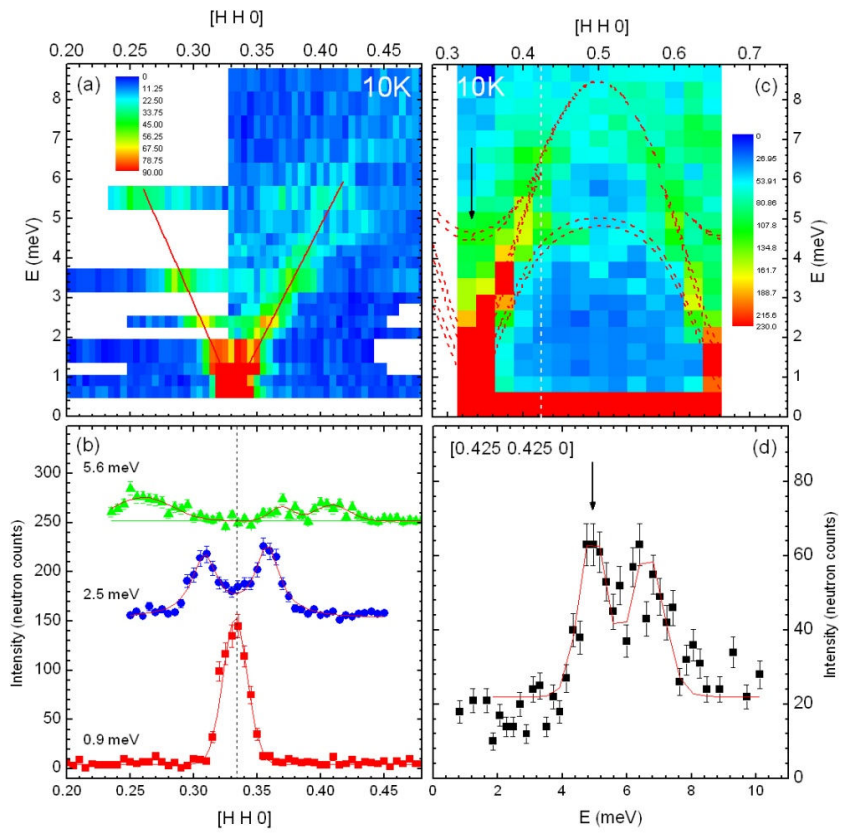

Figure 5

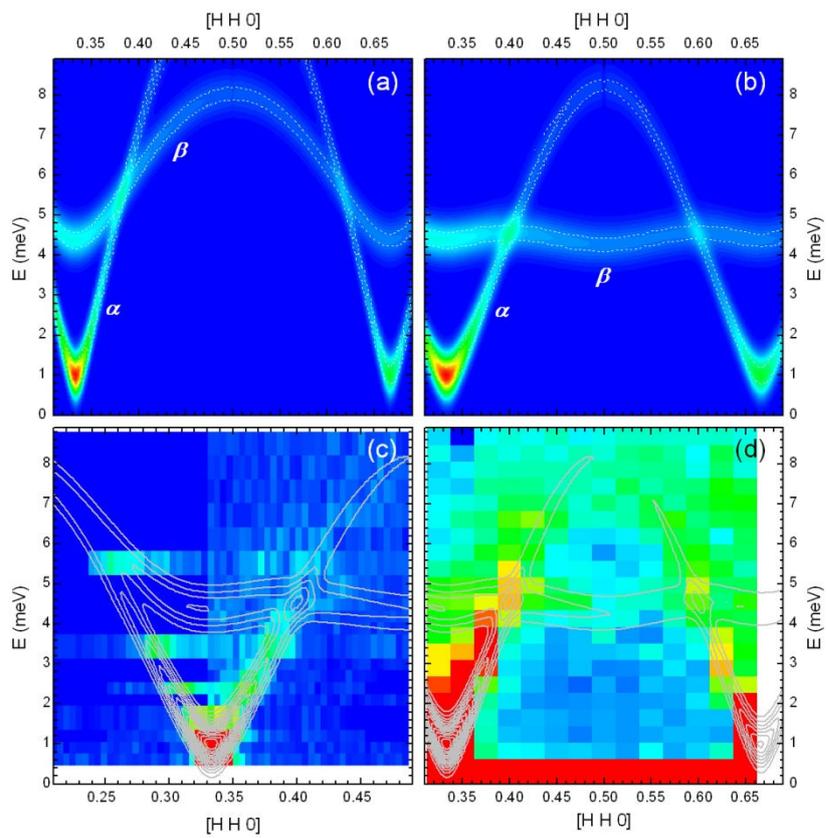

Figure 6

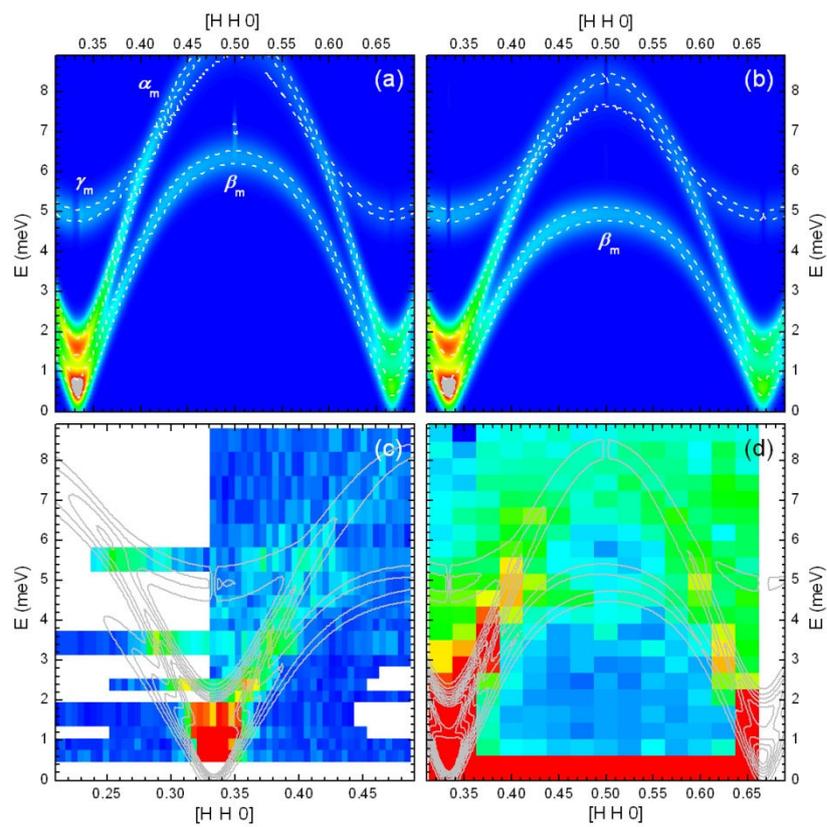

Figure 7

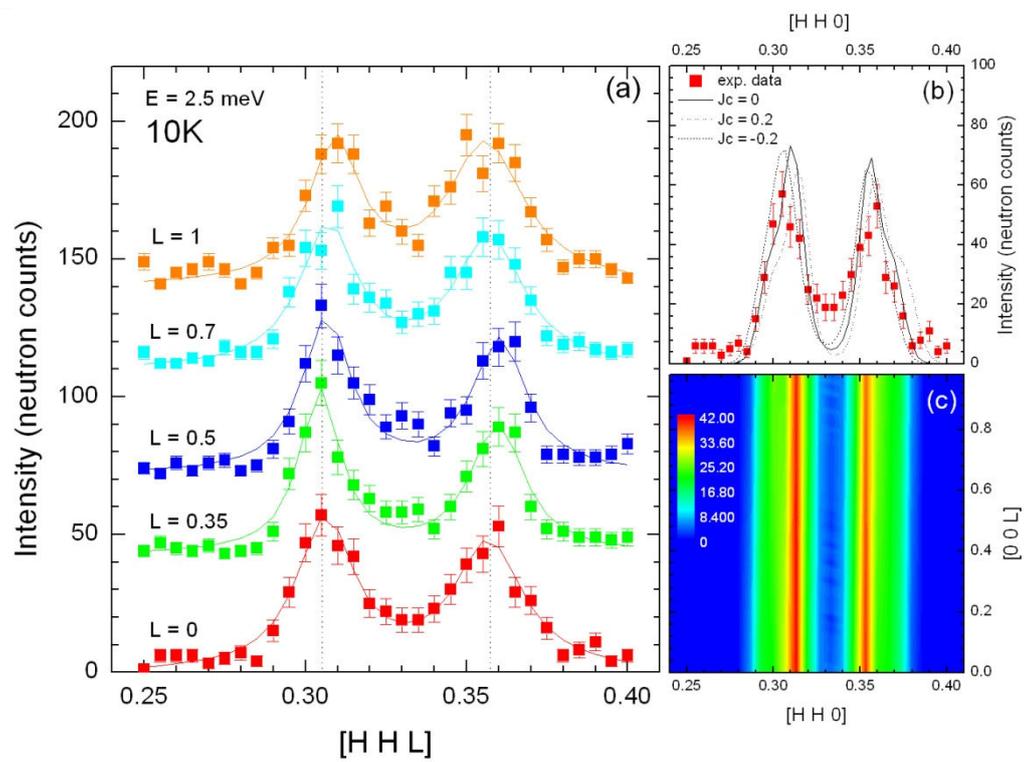